\documentclass[pre,floatfix]{revtex4-1} 
\usepackage[pdfpagemode=UseNone,pdfstartview=FitH,colorlinks=true]{hyperref}
\usepackage{enumitem}
\usepackage{url}
\setlistdepth{9}
\usepackage{graphicx}
\usepackage{epstopdf}

\begin{document}
\title{Associative memory by collective regulation of non-coding RNA}

\author{J. M. Deutsch}
\email{josh@ucsc.edu}
\affiliation{Department of Physics, University of California, Santa Cruz CA 95064}

\begin{abstract}
The majority of mammalian genomic transcripts do not directly code for proteins
and it is currently believed that most of these are not under evolutionary constraint. However given the
abundance non-coding RNA (ncRNA) and its strong affinity for inter-RNA binding, these molecules have the potential
to regulate proteins in a highly distributed way, similar to artificial neural
networks. We explore this analogy by devising a simple architecture for a biochemical network that can  
function as an associative memory. 
We show that the steady state solution for this chemical network has the same structure as an associative 
memory neural network model. 
By allowing the choice of equilibrium constants between
different ncRNA species, the concentration of unbound ncRNA can be made to follow any pattern and many 
patterns can be stored simultaneously.  The model is studied numerically
and within certain parameter regimes it functions as predicted.
Even if the starting concentration pattern is quite different, it is
shown to converge to the original pattern most of the time. The network is also
robust to mutations in equilibrium constants. This calls into
question the criteria for deciding if a sequence is under evolutionary constraint.
\end{abstract}
\maketitle
 
\section{Introduction}

Non-coding RNA (ncRNA) has emerged in recent years as a major player in the molecular
biology of the cell. The vast majority of transcripts in the cell produce long
(greater than 200 nucleotides) non-coding RNA, and many of their functions have
been studied~\cite{ENCODEpilot,Kapranov2007,MercerDingerMattick,ENCODE}. A recent
paper~\cite{Deutsch14} proposed that a function of much of the non-coding
RNA thought not to be under evolutionary constraint, was to act collectively to
regulate protein transcription. Instead of individual genes being controlled by
a few regulatory elements, ncRNA acts collectively to control transcription. If
we assume weak binding of many species of ncRNA to each other, the amount of
unbound ncRNA depends on all of the other species collectively. This free ncRNA
can then act to regulate mRNA in a variety of possible ways. The important 
distinction between this behavior and other regulatory mechanisms, such as  {\em cis}-regulation,
is the collective nature of regulation.

In computer science, it has been known for many decades, that it is possible to make intelligent
computational decisions using {\em distributed} networks, and these form the
bulk of models for artificial and real neural network~\cite{HertzKroghPalmer}. This differs from
most digital circuitry in that the output is computed
using network elements that interact with many others in a robust way. This
means that even if connection strengths are varied or even eliminated entirely, the network will 
still keep much of its original function. This is an advantage over traditional
computer circuitry which has a much sparser connectivity. Furthermore this
architecture is well suited to learning new tasks.

In this paper, we devise a ncRNA system that will behave as an {\em associative memory}. The model behaves in a 
similar manner to associative memory neural networks~\cite{LittleShaw,Hopfield,AmitGutFreundSompolinsky}.
There are two ingredients to this system: {(\em i)} A regulated mechanism to produce many
species of ncRNA molecules, that will degrade at some rate over time, {(\em ii)}
The ncRNA molecules bind and unbind with equilibrium constants that can be
varied by changing the sequence of them. We assume that the regulation will
depend on the total concentration of ncRNA, and the amount of unbound ncRNA. We
find a specific functional form for the regulation of ncRNA that leads to
nonlinear self-consistent equations for unbound ncRNA concentrations. These
equations are essentially the same as what is used in associative memory models. 

There are three caveats to this scenario however. The first is that the
relaxation time for the chemical equilibration of the ncRNA is much shorter 
than the time it takes to degrade ncRNA
molecules. The second, is that we have given the mathematical form for the
regulation of ncRNA that depends on different ncRNA concentrations, but we have
not provided a physical or chemical model that can implement this precisely.
Similar regulation takes place but its precise form is still not understood well
quantitatively. Finally, it requires that we have complete freedom to choose the
equilibrium constants between different species. 

As with many neural network and statistical mechanical models, the
precise form of a model often does not matter~\cite{HertzKroghPalmer}, and we expect that a range of models
will have similar behavior. This model does not imply that precisely
this biochemical mechanism exists, but rather it shows the existence of
mechanisms that are relatively simple, and can result in collective regulation
of the genome in a manner quite unlike the ones that have already been discovered. 
This work makes such a paradigm more plausible.

\section{The model}

Here we study how regulating the transcription of $N$ species of non-coding RNA (ncRNA), combined with the
promiscuous binding of ncRNA species to each other, can lead to an associative
memory. We denote the total concentration of of the $N$ species by $C_1, C_2, \dots, C_N$, and the unbound concentrations
$\rho_1, \rho_2,\dots,\rho_N$. 

\subsection{Model Assumptions}

\subsubsection{Assumptions about binding}

We assume that the standard form for chemical reactions between different
species, $i$, and $j$
\begin{equation}
i + j \rightleftharpoons ij
\end{equation}
We are assuming here that these are the only kinds of reaction present. There
are no higher order reactions involving three or more species.

We can define an equilibrium constant~\cite{Reif} $K_{i,j} = \rho_{ij}/\rho_i\rho_j$.
Here $\rho_{ij}$ is the concentration of bound molecules $ij$. This leads
to~\cite{Deutsch14}
\begin{equation}
\label{eq:ConcentrationModification}
\rho_i =\frac{C_i}{1+\sum_j\rho_j K_{i,j}}
\end{equation}
for $i=1,\dots,N$.

We further assume that we have complete freedom to choose the equilibrium
constants. That is, there is enough choice in the sequences and binding
positions of the RNA molecules that the equilibrium constant of one pair does
not influence another one.

\subsubsection{Assumptions about molecular regulation}

The concentration of an individual species of ncRNA needs to be regulated for the
mechanism described here to work. Regulation is expected for biochemical
processes, however we require more specifically that the regulation should depend on
the current total concentration $C$, {\em as well as} the concentration of unbound ncRNA
$\rho$. $\rho$ can be measured with the assumption that only the unbound
molecules will be able to bind to a particular RNA binding biomolecule (such as
a protein) involved in the regulation of this kind of ncRNA. The total concentration $C$, can be measured
with the assumption that there is another biomolecule that binds to a different
portion on the ncRNA which is normally not associated with another RNA molecule,
for example because it is in a stem loop. An example of such a protein is a
stem-loop binding protein (SLBP). 

If both unbound and bound ncRNA concentrations are directly related to the concentrations of certain proteins
that bind to them as suggested above, 
it would appear feasible that a complex of such proteins could evolve to
produce transcription machinery that allows regulation in the way
we hypothesize below. This seems
reasonable because well known genetic regulatory mechanisms are well tuned to both enhance and suppress
transcription. We are only requiring that individual components
enhancing and suppressing transcription can be combined to mimic the smooth
dependence on $\rho$ and $C$ that is used here.

\subsection{Dynamics of concentration and transcription rates}

Eq. \ref{eq:ConcentrationModification} relates the unbound concentration of one
species to its total concentration, and the unbound concentration of all the
other species. The system will start off out of equilibrium and we assume that
it approaches it with a simplified first order kinetic equation
\begin{equation}
\label{eq:FirstOrderRho}
\tau_\rho\frac{d\rho_i}{dt} = -\rho_i + \frac{C_i}{1+\sum_j\rho_j K_{i,j}}
\end{equation}
for $i=1,\dots,N$,
where $\tau_\rho$ is a relaxation time, giving the time scale for the relaxation of
the $\rho_i$'s to equilibrium.

As discussed above, the concentrations $C_i$ are regulated. We assume that
individual molecules degrade over a timescale $\tau_C$, but are also being
transcribed, leading to a non-zero steady state concentration. We assume that
the degradation is a first order process, and the transcription of new ncRNA
molecules is a function that depends on $C_i$ and all of the $\rho$'s.  
\begin{equation}
\label{eq:FirstOrderC}
\tau_C \frac{d C_i}{dt} = - C_i + f(C_i,\{\rho_k\}) 
\end{equation}
Physically we expect that $\tau_C \gg \tau_\rho$, that is, the degradation of
the RNA happens at a much slower rate than equilibration of the binding and
unbinding of RNA.  $f$ represents the rate at which new transcripts are being
produced. This could be accomplished by regulating the transcription through
a variety of means, such as repressors or coactivators that can sense unbound
and bound concentrations as discussed above.

We now choose a specific form of the function $f$ that will lead to an
associative biochemical memory.
\begin{equation}
\label{eq:FormOfF}
f(C_i,\{\rho_k\}) = \frac{C_i}{\rho_i} S(4\frac{C_i}{\rho_i} - 2\sum_j (K_{i,j} + \rho_j) -3N)
\end{equation}
for $i=1,\dots,N$.
The function $S(x)$ is a sigmoid function that could take a variety of forms,
such as a logit function
\begin{equation}
\label{eq:Sigmoid}
S(x) = \frac{1}{2}[1+\tanh(\beta x/N)]
\end{equation}
where $\beta$ is a constant, that in analogy to spin systems, represents an
inverse temperature. We shall discuss its value below when we discuss the
numerical implementation of this model.

Eqs. \ref{eq:FirstOrderRho}, \ref{eq:FirstOrderC}, \ref{eq:FormOfF}, and \ref{eq:Sigmoid} define
the dynamics of the $2N$ concentrations given a set of equilibrium constants
$K_{i,j}$, and the initial conditions.

\section{Analysis}

In steady state, all time derivatives vanish, and Eqs \ref{eq:FirstOrderRho} 
reduces to \ref{eq:ConcentrationModification}. It can be rearranged to yield:
\begin{equation}
\label{eq:KP}
\sum_j K_{i,j} \rho_j = \frac{C_i}{\rho_i}-1
\end{equation}
Eq. \ref{eq:FirstOrderC} reduces to
\begin{equation}
\label{eq:SelfConsistentC}
C_i = f(C_i,\{\rho_k\}) 
\end{equation}
If we transform to symmetric variables $s_i$, $i=1,\dots,N$ so that $\rho_i = (1+s_i)/2$. 
We will also choose symmetrized variables $J_{i,j}$ so that 
\begin{equation}
\label{eq:JtoK}
K_{i,j} = \frac{1+J_{i,j}}{2}.
\end{equation}
Substituting this into Eq. \ref{eq:FormOfF} and using Eq. \ref{eq:KP} gives
\begin{equation}
\label{eq:FTransformed}
C_i = f(C_i,\{s_k\}) = \frac{C_i}{1+s_i/2} S(\frac{1}{N}\sum_j J_{i,j} s_j)
\end{equation}
Canceling the $C_i$'s solving for $s_i$, and substituting Eq. \ref{eq:Sigmoid}
finally gives
\begin{equation}
\label{eq:Hopfield}
s_i = \tanh(\frac{\beta}{N}\sum_j J_{i,j} s_j)
\end{equation}

Note that there is no guarantee that the dynamics of this system will lead to
these steady state solutions as some of these may be unstable. But for the physically
sensible requirement that we have imposed on the relaxation times, we will
see below that these solutions are achieved. 

This self consistent equation is used in analyzing neural network associative
memories~\cite{LittleShaw,Hopfield,AmitGutFreundSompolinsky} and can have many
solutions, depending on the choice of the $J$'s. We will now focus on the most
common choice, that of Hebbian learning~\cite{Hebb,LittleShaw,Hopfield}. 

\subsection{Hebbian coupling}

We can choose the $K$'s or equivalently the $J$'s in a manner that will allow
the network to retrieve $M$ patterns. The retrieval works best when the patterns
are uncorrelated. Let us denote the $\alpha$th pattern by $t_i^\alpha$,
$\alpha = 1,\dots,M$. We let each $t_i^\alpha = \pm 1$. Then 
\begin{equation}
\label{eq:Hebb}
J_{i,j} = \frac{1}{M} \sum_{\alpha=1}^M t_i^\alpha t_j^\alpha
\end{equation}
The number of patterns that can be reliably stored is proportional to
$N$~\cite{AmitGutFreundSompolinsky} but
this will also depend on the correlations between the patterns.

In order for this choice to be physically meaningful, the $K$'s must all be
positive. This is why we chose to relate the $K$'s to the $J$'s in the first
place through Eq. \ref{eq:JtoK}.
After choosing $M$ patterns and setting the $J$'s accordingly, they are shifted
and scaled to produce positive $K$'s. $K \rightarrow K-min(K)$, and $K \rightarrow K/max(K)$.
This produces equilibrium constants between zero and one.

\subsection{Numerical results}

The above model was implemented numerically. We chose $N=50$ ncRNA species, and 
analyze the retrieval $M=3$ separate patterns using the Hebbian rule of Eq. \ref{eq:Hebb}.

The ratio of the two timescales $\tau_C/\tau_\rho$ was found to be important in
the convergence of this system. With $\tau_C/\tau_\rho = 100$ the system
always appeared to converge. However with $\tau_C/\tau_\rho = 10$, it sometimes
did not. 

\begin{figure}[htb]
\begin{center}
\includegraphics[width=0.7\hsize]{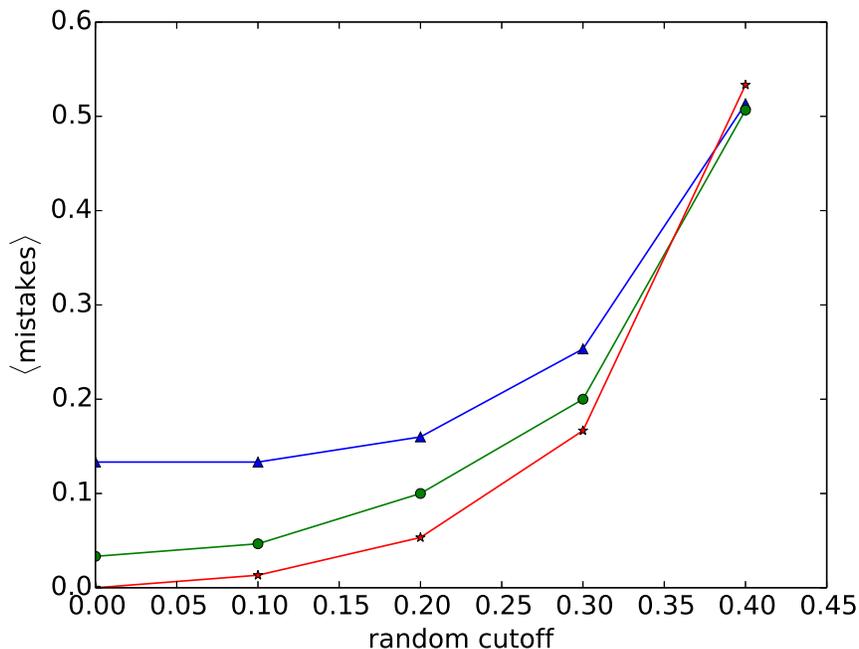}
\caption
{ 
The number of mistakes made as a function of the distance away from a pattern.
When the initial condition for the density is sufficiently different from the
pattern, the system will move away from it towards a different to a different
pattern. The red curve (stars) shows $\beta = 8$, the green (circles), $\beta = 7$, and
the blue (triangles) $\beta = 6$. The lines are a guide for the eye.
}
\label{fig:mistakesvscutoff}
\end{center}
\end{figure}

We tested out the basin of attraction of initial values of the $\rho_i$'s. We
considered each pattern $t^\alpha$ and then randomly altered its sequence by
varying amounts. For each ncRNA species, we chose a random number between zero and 1 flipped the value
of the density $\rho_i \rightarrow 1-\rho_i$, if the random number was below
some cutoff. We tried $5$ different mutations for each of the $M$ patterns, for
a fixed cutoff. We generated $10$ separate examples of the $M$ patterns, and
varied the cutoff. We plot the fraction of mistakes as a function of this
cutoff in Fig \ref{fig:mistakesvscutoff}. The three graphs represents three
different values of $\beta$: $6$, $7$, and $8$. It is evident that for $\beta$
equals $6$, or $7$, some patterns are not stable, because even with a cutoff of
zero leads in some cases to the system moving away to a different pattern.
However, with $\beta =8$, all patterns were stable.

\begin{figure}[htb]
\begin{center}
\includegraphics[width=0.7\hsize]{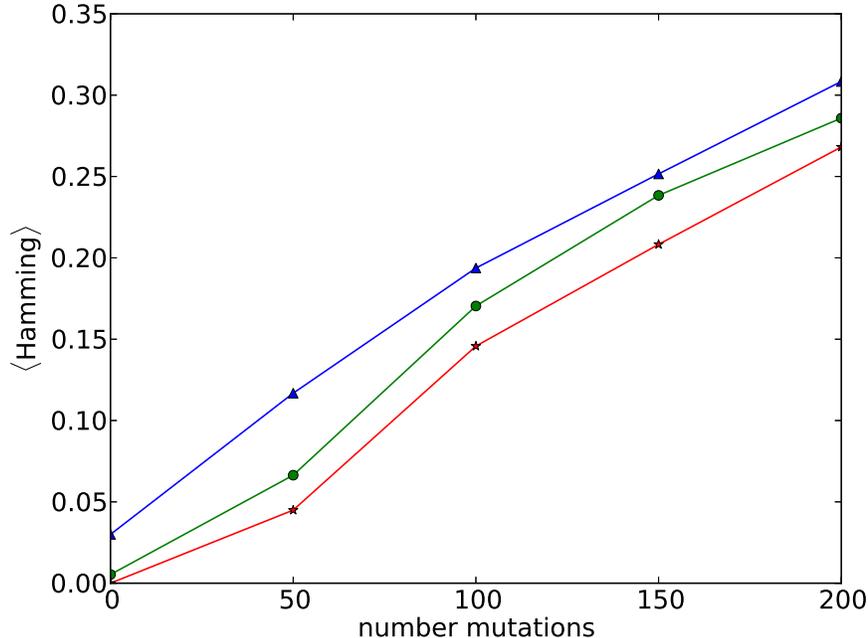}
\caption
{ 
The effects of mutations to the equilibrium constants. The horizontal axis gives
the number of mutations that were made. The vertical axis is the average fraction of errors, measured by the
fractional Hamming distance of the desired pattern to the recalled pattern. This
represents the fraction of incorrect concentration levels.
The red curve (stars) shows $\beta = 8$, the green (circles), $\beta = 7$, and
the blue (triangles) $\beta = 6$. The lines are a guide for the eye.
}
\label{fig:hamming}.
\end{center}
\end{figure}

Next we tested out the effects of mutating the equilibrium constants $K_{i,j}$. We chose
random $i$'s and $j's$ and mutated them by taking $K_{i,j} \rightarrow 1 - K_{i,j}$,
maintaining symmetry of the matrix. (Recall that
$0 <= K_{i,j} <= 1$). We measured how close the final fixed point was to the
original pattern, by measuring the Hamming distance between the two and dividing
by $N$.  We varied the number of $K$'s that were mutated. We ran this for the
same number of times as above, and for the same values of $\beta$. The results
are shown in Fig. \ref{fig:hamming}. Even with $100$ mutations, most of a
pattern is correctly recalled. This illustrates that this architecture is robust
to mutations.

\section{Discussion}

Here we have developed and analyzed a model for ncRNA that produces features of
an associative memory which assumes promiscuous binding of many species to each other. 
We mapped the fixed points of this biochemical network onto a neural
network model. However the stability of these points
depends on two relaxation times, one for ncRNA equilibration, and the
other characterizing the degradation time of ncRNA molecules. Furthermore the switching on
of ncRNA transcription must be a sharp enough function of ncRNA concentrations, 
that is, have high enough $\beta$ (see Eq. \ref{eq:Sigmoid}), for the fixed points to be stable.

Although an associative memory has been considered here mainly as a gross simplification
of a real genetic network, in order to elucidate, in detail,
how genetic networks could make use of collective regulation, it is possible that this kind of
behavior could be useful in an organism. In a neural network, if some fraction of a pattern is shown
to an associative memory, the rest of it will be reproduced. Similarly for a
genetic network, if some ncRNA species are presented to the network at given
concentrations, the network will recognize this sub-``pattern", and produce the
additional pattern of ncRNA molecules at desired concentrations. Multiple 
patterns can be stored, meaning that the network can respond appropriately to different
environmental conditions.

There are many advantages to distributed network computation and one would
expect, as argued here, that these would carry over to networks used for genetic regulation.
The robustness to mutation is one such feature. This means that mutations of
ncRNA will have a much gentler effect than for sparse networks, but nevertheless
could confer evolutionary advantages. This is important
because much ncRNA is thought not to be under evolutionary constraint
but in light of the above, this may be an incorrect conclusion.
Much larger mutation rates for molecules utilizing collective regulation are
expected even if they are involved in important regulatory functions. 
Therefore the mutation rate criteria for evolutionary pressure 
should be questioned to take this kind of collective mechanism.

The robustness of collective regulations to changes also means that it is difficult to come
up with experimental means to find it, as each pairwise interactions between
species are small and removal of individual interactions will not be easily
noticed.

This model was not evolutionary, as this would make analysis intractable
analytically. However it is straightforward to see how evolution would give rise
to equilibrium constants that could store multiple patterns. Indeed, this is 
the similar to what is done with ``Boltzmann machines"~\cite{HertzKroghPalmer},
where the network is trained with Monte Carlo on multiple patterns and eventually learns to
respond to all of them correctly. The couplings will not end up being precisely
those of Eq. \ref{eq:Hebb}, but they will achieve the same goal. Therefore we expect
that an architecture such as described here, that evolves by mutating the
$K_{i,j}$'s will produce similar behavior.

This work was supported by the Foundational Questions Institute \url{<http://fqxi.org>}.

\end{document}